\documentclass[aps,superscriptaddress,twocolumn,preprintnumbers,floatfix,showpacs,prx]{revtex4-1}

\usepackage{amsmath,amssymb}
\usepackage{dcolumn}
\usepackage{braket}
\usepackage{dsfont}
\usepackage{feynmp}
\usepackage{hhline}
\usepackage{graphicx}
\usepackage{mathtools}
\usepackage{xfrac}
\usepackage{paralist}
\usepackage{enumitem}
\usepackage[normalem]{ulem}
\usepackage{color}
\usepackage[draft]{pgf}

\definecolor{darkblue}{rgb}{0.,0.,0.4}
\definecolor{darkred}{rgb}{0.5,0.,0.}

\newcommand{\bk}{{\bf k}}
\newcommand{\bK}{{\bf K}}
\newcommand{\bq}{{\bf q}}
\newcommand{\bp}{{\bf p}}

\newcommand{\beq}{\begin{eqnarray}}
\newcommand{\eeq}{\end{eqnarray}}

\newcommand{\beqq}{\begin{eqnarray*}}
\newcommand{\eeqq}{\end{eqnarray*}}

\newcommand{\be}{\begin{equation}}
\newcommand{\ee}{\end{equation}}

\begin{document}

\title{Non-Hermitian Topological  Theory of Finite-Lifetime Quasiparticles: \\ Prediction of Bulk Fermi Arc Due to Exceptional Point}
\author{Vladyslav Kozii}
\author{Liang Fu}
\affiliation{Department of Physics, Massachusetts Institute of Technology, Cambridge, MA 02139, USA}
\begin{abstract}
We introduce a topological theory to study quasiparticles in interacting and/or disordered many-body systems, which have a finite lifetime due to inelastic and/or elastic scattering. The one-body quasiparticle Hamiltonian includes both the Bloch Hamiltonian of band theory and the self-energy due to interactions, which is non-Hermitian when quasiparticle lifetime is finite. We study the topology of non-Hermitian quasiparticle Hamiltonians in momentum space, whose energy spectrum is complex. The interplay of band structure and quasiparticle lifetime is found to have remarkable consequences in zero- and small-gap systems. In particular, we predict the existence of topological exceptional point and bulk Fermi arc in Dirac materials with two distinct quasiparticle lifetimes.

\end{abstract}
\maketitle

The intensive theoretical and experimental study of topological materials over the past decade has made the notion of topology an important and useful concept in condensed matter physics. In a very broad sense, topology is concerned with robust properties of the system that remain unchanged upon perturbations. Since the pioneering work of Thouless-Kohmoto-Nightingale-den Nijs \cite{TKNN} on quantum Hall systems, much attention has been focused on the topological characterization of quantum states of matter at zero temperature. A quantum state is called topological if its ground state wavefunction bears a distinctive character that can be captured by a topological invariant.
A recent example of topological quantum states is topological insulator \cite{HasanKane}, whose realization has been found in numerous materials \cite{Ando}.

Despite its enormous success, the current use of topology has its limitations in studying quantum many-body systems at finite temperature. From a conceptual perspective, the strict distinction between topological and trivial insulators is lost at $T\neq 0$, as the non-analyticity of free energy in the gap-closing transition at $T=0$ is rounded.
Thus topological characterization of $T=0$ quantum phases is only  useful at temperatures much smaller than the gap, where the correction to quantized physical observables such as Hall conductance is small.


Correlated electron systems are known to exhibit a wide variety of complex phenomena at finite temperature, even when $T=0$ ground states are relatively simple. The most notable example is the strange metal and pseudogap phase of cuprates above the superconducting transition temperature. This motivates us to consider whether the study of correlated electron systems at finite temperature may benefit from topology broadly defined, rather than in the restricted sense of topological quantum states.

In this Letter, we introduce a topological theory of finite-lifetime quasiparticles.  
Finite quasiparticle lifetime is a generic property of realistic quantum many-body systems, resulting from either inelastic electron-electron/electron-phonon scattering at $T \neq 0$, or elastic electron-impurity scattering.
We find that the interplay of band structure and quasiparticle scattering has remarkable effects on quasiparticle dispersion and dynamics in zero- or small-gap systems in general. In particular, we predict that quasiparticle lifetime effect completely reshapes the low-energy dispersion of Dirac materials, leading to a topologically protected bulk Fermi arc. 

The central object of our theory is the quasiparticle Hamiltonian $H$ which we {\it define} from retarded electron Green's function $G^R$,
\beq
G^R(\omega) &=& (\omega - H(\bk,\omega) )^{-1}, \label{GR} \\
H(\bk, \omega)& \equiv & H_0(\bk) + \Sigma (\bk, \omega). \label{HG}
\eeq
where $H_0(\bk)$ is the single-particle Hamiltonian of Bloch electron in the periodic potential, and $\Sigma(\bk, \omega)$ is electron's self-energy that includes the effect of electron-electron, electron-phonon and electron-impurity scatterings. 
Importantly, while the Bloch Hamiltonian $H_0$ is Hermitian, the self energy $\Sigma$ is {\it non-Hermitian} when quasiparticle lifetime is finite.
Then the quasiparticle Hamiltonian  $H(\bk, \omega)$ is also non-Hermitian and its energy spectrum ${\cal E}_n(\bk, \omega)$ is complex, where $n$ is the band index.

At a given $\bk$, the spectrum of $H(\bk, \omega)$, i.e., ${\cal E}_n(\bk, \omega)$,  can have a discrete set of poles $\omega = z_n(\bk) $ in electron Green's function in the complex frequency plane, such that ${\cal E}_n(\bk, z_n)= z_n$. In the vicinity of a first-order pole, the Green's function takes the form
$
G^R(\bk, \omega) \sim \frac{1}{\omega - z_n}
$. Then the real part of $z_n(\bk)$ defines the quasiparticle energy dispersion, while its imaginary part gives the quasiparticle inverse lifetime.

The quasiparticle Hamiltonian $H(\bk, \omega)$ thus provides a natural generalization of Bloch Hamiltonian $H_0(\bk)$ for noninteracting electrons, and encodes a wealth of information about quasiparticles dispersion and dynamics.
The goal of this work is to study the topological property of $H(\bk, \omega)$ exclusive to its non-Hermitian nature, and to explore the implication of topology for the behavior of finite-lifetime quasiparticles. %


A fundamental difference between Hermitian and non-Hermitian Hamiltonians is that the latter can be non-diagonalizable, i.e., its linearly independent eigenstates do not span the full Hilbert space. This leads to the possibility that the quasiparticle Hamiltonian $H(\bk,\omega)$ becomes non-diagonalizable at certain momentum, called ``exceptional points'' in the mathematical physics literature \cite{Kato}.
Interestingly, exceptional points in two and higher dimensions are topologically stable and characterized by a nontrivial topological index \cite{ShenZhenFu}. In recent years, there has been a growing interest in exceptional points in open systems \cite{EP-review,BoZhen}, for which the Hamiltonian is intrinsically non-Hermitian due to the coupling with an external bath. In contrast, here we are interested in interacting or disordered systems, for which the microscopic many-body Hamiltonian is Hermitian, but the one-body quasiparticle Hamiltonian $H(\bk,\omega)$ is non-Hermitian due to the finite quasiparticle lifetime.  In the presence of electron-electron or electron-phonon interaction,
a quasiparticle decays into incoherent multi-particle excitations, while in the presence of disorder, elastic scattering leads to a finite lifetime for a quasiparticle of a given momentum.




We now demonstrate the emergence and consequence of topological exceptional points in the quasiparticle spectrum of zero- and small-gap materials. Before considering the general case, we first study a model of Dirac semimetals in two dimensions, where the Dirac dispersion comes from two different orbitals unrelated by symmetry. This scenario may be applicable to heavy fermion systems \cite{Coleman}, where the two orbitals correspond to itinerant and localized electrons and the hybridization gap can have a $p$-wave form factor leading to Dirac point nodes.

The Bloch Hamiltonian of our two-orbital model is given by
\beq
H_0(\bp) = \left(
\begin{array}{cc}
\epsilon_1 + \frac{\bp^2}{2m_1} & v_y p_y \\
v_y p_y & \epsilon_2  -\frac{\bp^2}{2m_2}
\end{array}
\right) \label{H0}
\eeq
where the diagonal part describes the electron- and hole-like dispersion of the two orbitals, and the off-diagonal part describes their hybridization. Here we assume the two orbitals are even and odd under both spatial inversion and the reflection $y\rightarrow -y$, respectively.  Under this condition, the hybridization term must be an odd function of $p_y$, and $H_0(\bp)$ must be real due to the presence of both inversion and time-reversal symmetry.

In our model, a pair of Dirac points appear at opposite momenta when $\epsilon_1 < \epsilon_2$, i.e., the band gap at $\bp=0$ is inverted.   Linearizing the Hamiltonian $H_0(\bp)$  near a Dirac point located at $\bp_0=( \sqrt{2m_1 m_2 (\epsilon_2 - \epsilon_1)/(m_1 + m_2)} ,0)$ yields the low-energy Hamiltonian:
\beq
H_0(\bk) = \left(
\begin{array}{cc}
 v_1 k_x & v_y k_y \\
v_y k_y & -  v_2 k_x
\end{array}
\right) \label{Heff}
\eeq
with $\bk=\bp - \bp_0$.
In general, $v_1 \neq v_2$ so that the Dirac cone is tilted in the $k_x$ direction.  This tilting is unimportant for this work, and for simplicity we set $v_1 = v_2 =v_x$ below.

Despite its simplicity, our model is motivated by general symmetry considerations and  applies to several known materials. The two-orbital Hamiltonian (\ref{H0}) captures the band inversion between cation $d$-orbitals and anion $p$-orbitals of monolayer transition metal dichalcogenides in the 1T' structure, when the small spin-orbit gap at the Dirac point is neglected \cite{Qian}. The Dirac Hamiltonian (\ref{Heff}) also applies to surface states of topological crystalline insulators (Pb,Sn)Se, where the two orbitals come primarily from the Pb/Sn and Se atoms respectively \cite{TCI-Madhavan}.

Now consider the general form of electron self-energy $\Sigma(\bk, \omega)$ in our model, which may come from electron-phonon, electron-electron, and/or electron-impurity scattering. Provided that electron interaction does not lead to spontaneously symmetry breaking, $\Sigma(\bk, \omega)$ respects the reflection symmetry as the Dirac Hamiltonian (\ref{Heff}) does, i.e.,
\beq
\Sigma(k_x, k_y, \omega) =\sigma_z \Sigma(k_x, -k_y,  \omega) \sigma_z, \label{reflection}
\eeq
where $\sigma_z = \pm 1$ denotes the two orbitals. In general, $\Sigma(\bk, \omega)$ is the sum of a Hermitian part $\Sigma'(\bk, \omega)$ and an anti-Hermitian part $\Sigma''(\bk, \omega)$: $\Sigma = \Sigma' + i \Sigma''$.  $\Sigma'$ renormalizes the bare band structure given by the Bloch Hamiltonian (\ref{Heff}), while $\Sigma''$ leads to a finite quasiparticle lifetime, which we shall focus on below.

We are mainly interested in self-energy on quasiparticles near the Dirac point, where its effect is most prominent. If $\Sigma''(\bk, \omega)$ is an analytic function of $\bk$, it can be expanded in powers of $\bk$. To leading order, $\Sigma''(\bk, \omega)\sim \Sigma''(0,\omega)$ must be a diagonal matrix due to the reflection symmetry (\ref{reflection})
\beq
i \Sigma''(\bk,\omega ) \simeq   \left(
\begin{array}{cc}
 i \Gamma_1(\omega) & 0 \\
0 &  i\Gamma_2(\omega)
\end{array}
\right), \label{lifetime}
\eeq
where $\Gamma_{1,2}$ is the inverse lifetime of the two orbitals respectively. Since these two orbitals are assumed to be unrelated, we expect $\Gamma_1\neq \Gamma_2$ in general. The presence of two distinct lifetimes will play a crucial role below.

As a concrete example, we study the self-energy due to electron-phonon scattering, leaving the cases of electron-electron and electron-impurity scattering to separate works. The full electron-phonon Hamiltonian of Dirac semimetal takes the form
$
H = H_{\text{el}} + H_{\text{ph}} + H_{\text{el-ph}}.
$
The first term is the electronic Hamiltonian of low-energy Dirac fermions at chemical potential $\mu$,
$
H_{\text{el}} =\sum_{\bk} c_\bk^\dagger (H_0(\bk) - \mu)  c_\bk
$
where $c^\dagger_\bk = (c^\dagger_{1\bk}, c^\dagger_{2\bk})$ with $i=1,2$ denoting the two orbitals.
$
H_{\text{ph}}=\omega_0 \sum_{\bq} \left( b^\dagger_\bq b_\bq + \frac12  \right)
$ describes dispersionless Einstein phonons at frequency $\omega_0$,
where $b_\bq^\dagger$ is the phonon creation operator. The electron-phonon interaction is given by
\be
H_{\text{el-ph}} =  \sum_{\bk,\, \bq} (\lambda_1 c_{1\bk + \bq}^\dagger c_{1\bk} + \lambda_2 c_{2\bk + \bq}^\dagger c_{2\bk })\phi_{\bq}
\ee
where $\phi_\bq = (b_\bq + b^\dagger_\bq)/\sqrt{2}$ is the lattice displacement operator, and $\lambda_{i}$ is the electron-phonon coupling constant for orbital $i$.

At weak coupling, we calculate the electron self-energy to lowest order in $\lambda_i$, and find  $\Sigma''(\bk,\omega )$  has the exact form of Eq.(\ref{lifetime}) with two inverse lifetimes $\Gamma_i$ given by
\begin{widetext}
\beq
\Gamma_{i}(\omega) &=& \frac{\lambda_i^2}{16 v_x v_y} \left( |\omega_0 + \mu +\omega| (\tanh \frac{\omega_0 + \omega}{2T} -\coth \frac{\omega_0}{2T} )
+ |\omega_0 - \mu - \omega| (\tanh \frac{\omega_0 - \omega}{2T} -\coth \frac{\omega_0}{2T} )  \right).
\eeq
\end{widetext}
At small $\omega$, $\Gamma_i(\omega) \simeq \Gamma_i(0)$ is a nonzero constant at any $T\neq 0$, hereafter denoted by $\Gamma_i$. 
For stronger electron-phonon coupling, a self-consistent calculation of self-energy is required. Nonetheless, our analysis below only relies on a nonzero $\Sigma''(\omega)$ in the limit of $\omega \rightarrow 0$, which we expect to hold in general at finite temperature.

After incorporating  the self-energy (\ref{lifetime}) into the Dirac Hamiltonian (\ref{Heff}), the $2\times 2$ quasiparticle Hamiltonian
$
H(\bk, \omega) = H_0(\bk) + \Sigma(\bk,\omega)
$ at small $\omega$ is given by
\beq
H(\bk)&=&  (v_x k_x - i \gamma) \sigma_z + v_y k_y \sigma_x - i \Gamma,  \label{Hg}
\eeq
with $\Gamma \equiv (\Gamma_1 + \Gamma_2)/2$ and $\gamma \equiv (\Gamma_1 - \Gamma_2)/2$. 
Note that the self-energy term $-i \gamma \sigma_z$ generally does not commute with the Bloch Hamiltonian.
The two quasiparticle bands of $H(\bk)$ have a complex-energy dispersion given by
\beq
{\cal E}_\pm(\bk) =  \pm \sqrt{v_x^2 k_x^2 + v_y^2 k_y^2 - \gamma^2 - 2i k_x \gamma} - i \Gamma,
\label{ek}
\eeq

 In the single-lifetime limit $\Gamma_1=\Gamma_2$ or $\gamma=0$, the real part of ${\cal E}_\pm(\bk)$, i.e., the quasiparticle energy, simply gives the Dirac dispersion $E_\pm(\bk)= \pm \sqrt{v_x^2 k_x^2 + v_y^2 k_y^2}$, while the imaginary part $-i\Gamma$ is simply a constant inverse lifetime for all states.

\begin{figure}[t]
\centering
\includegraphics[width=0.45\textwidth]{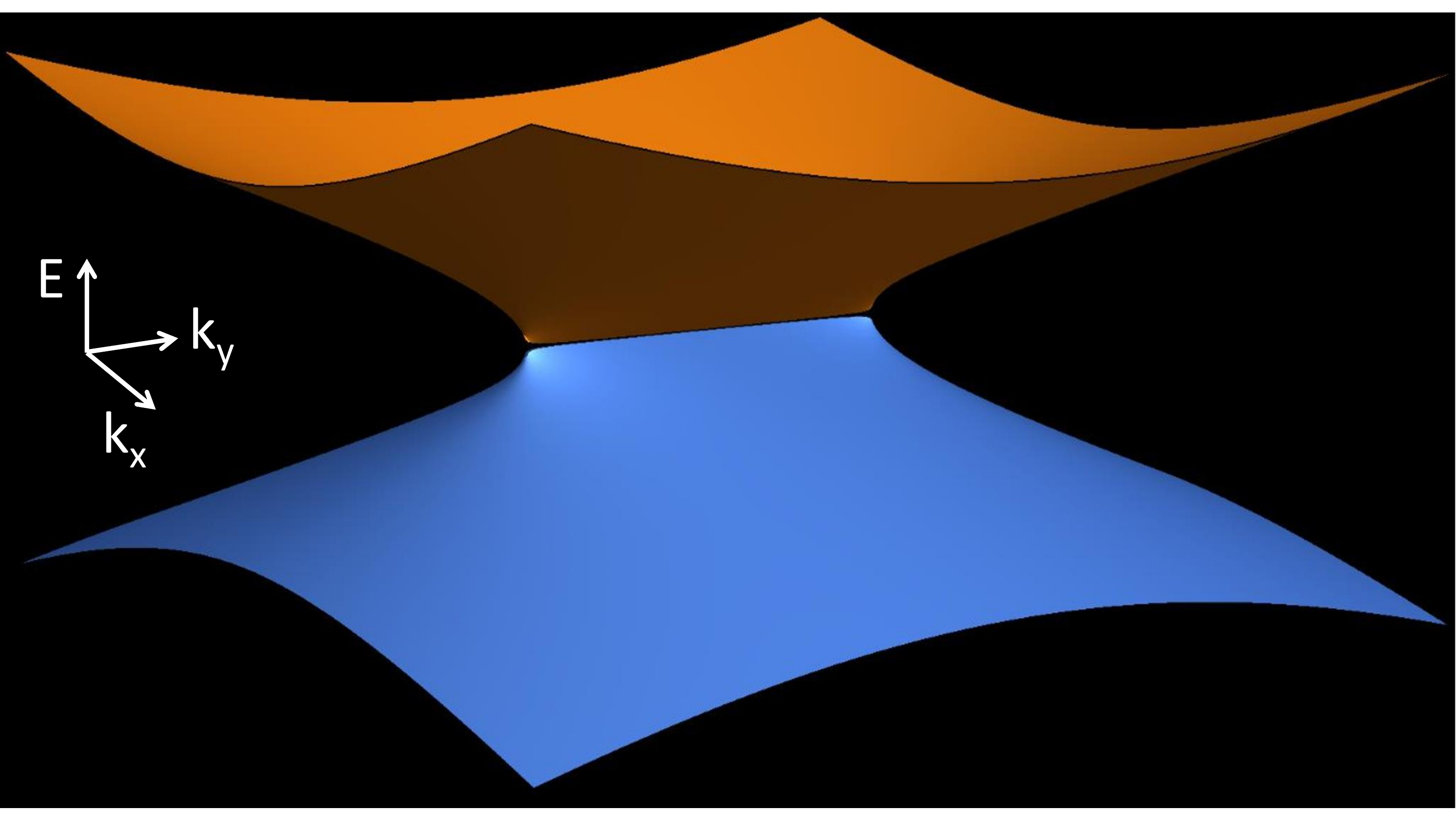}
 \caption{Quasiparticle energy dispersion of a two-dimensional Dirac semimetal reshaped by the two-lifetime self-energy, see Eq.(\ref{ek}). Instead of touching at the Dirac point, quasiparticle conduction and valence bands stick on a Fermi arc that ends at two topological exceptional points.}
\end{figure}

When the two lifetimes are unequal, i.e., $\gamma \neq 0$, $\cal E(\bk)$ in Eq.(\ref{ek}) is generally complex. The quasiparticle energy-momentum relation given by ${\rm Re} [ {\cal E}_\pm(\bk)]$, plotted in Fig.1, is dramatically different from the original Dirac dispersion at low energy. Instead of a single Dirac point at $\bk=0$, the quasiparticle conduction and valence bands touch on a line on the $k_y$ axis, where the dispersion is
\beq
{\cal E}_\pm (0,k_y) =  \pm \sqrt{v_y^2 k_y^2 - \gamma^2 } - i \Gamma
\eeq
For $|k_y|\leq \gamma/v_y \equiv k_0$, the real parts of both ${\cal E}_+$ and ${\cal E}_-$ are zero, leading to a line of band degeneracy that ends at two special momenta $(0, \pm k_0) \equiv \pm \bK$, which constitutes a bulk Fermi arc as we discuss in detail later. For $|k_y|> k_0$, a quasiparticle band gap $\Delta = {\rm Re} ({\cal E}_+ ) - {\rm Re} ({\cal E}_-)$ opens up. Near the end of the degeneracy line, the gap goes as
\beq
\Delta(0, k_y) = 2\sqrt{2 v_y \gamma (|k_y| - k_0) }, \; |k_y| > k_0
\eeq
rising up faster than near any degeneracy point in Hermitian band structures.

At the two end points, the quasiparticle Hamiltonian $H(\pm \bK)$ is  non-diagonalizable, as the $2\times 2$ matrix is defective---it only has a single eigenstate
$
\psi_{\pm \bK} = (1,\pm i)^T
$
with a purely imaginary eigenvalue $- i \Gamma$.  Near such a defective point $\pm \bK=(0,\pm k_0)$, the spectrum of $H(\pm \bK+\bq)$  is a double-valued holomorphic function of $q_x \pm i q_y$:
\beq
{\cal E} (\pm \bK+\bq) \simeq  \sqrt{2\gamma ( -iv_x q_x \pm v_y q_y)} - i \Gamma,  \label{ep}
\eeq
such that the two complex-energy bands are inseparable and form the Riemann surface for the square root.

Defective points with a square-root type spectrum (\ref{ep}) in the vicinity, are called exceptional points in band theory of non-Hermitian Hamiltonians \cite{ShenZhenFu}. Exceptional points are topologically stable and can only be created or annihilated in pairs. Their topological nature can be understood from the complex spectra of $H(\bk)$. As $\bk$ moves around a closed path enclosing an exceptional point, the two distinct eigenvalues ${\cal E}_+(\bk)$ and ${\cal E}_-(\bk)$ become swapped due to the square root singularity in (\ref{ek}), i.e., ${\cal E}_+ \rightarrow {\cal E}_-, {\cal E}_- \rightarrow {\cal E}_+$. This eigenvalue swapping defines a half-integer topological charge $\nu =\pm \frac{1}{2}$ associated with each exceptional point.

The swapping of complex eigenvalues ${\cal E}_+$ and ${\cal E}_-$ around an exceptional point dictates that the real parts of ${\cal E}_+$  and ${\cal E}_-$, or the quasiparticle energies of conduction and valence bands, must coincide at some $\bk$ on any closed path enclosing an exceptional point. Therefore, every  exceptional point is the end of a bulk Fermi arc where the quasiparticle bands are degenerate.

In our model, for any $\gamma \neq 0$ or equivalently $\Gamma_1 \neq \Gamma_2$, the original Dirac point splits into two exceptional points with opposite topological charges, which move further apart as $\gamma$ increases. The two exceptional points are connected by a band degeneracy line shown earlier.
Since the inverse lifetimes due to electron-phonon interaction increases with temperature, the locations of exceptional points in $\bk$ space and hence the length of the bulk Fermi arc are temperature dependent.

The presence of exceptional points also leads to a singular momentum dependence of quasiparticle lifetime, defined by the inverse of the imaginary part of $\cal E_\pm(\bk)$.  Across the Fermi arc, the quasiparticle lifetime of a given band has a discontinuous jump. This results from an abrupt change of quasiparticle wavefunction across the Fermi arc.

The topological nature of exceptional point does not rely on any symmetry, which makes it robust against all perturbations. Our model has shown that exceptional points appear in a zero-gap Dirac semimetal for any $\gamma\neq 0$. We now show that exceptional points and the concomitant bulk Fermi arcs are expected to be present in small-gap systems under broad conditions. This can be understood by considering the general form of the quasiparticle Hamiltonian $H(\bk,\omega=0)$ with two bands. It is convenient to express $H(\bk,\omega=0)$ as a $2\times 2$ matrix in the basis where the self energy $\Sigma''(\bk, \omega=0)$ is diagonal:
\beq
H(\bk,\omega=0)= \left(
\begin{array}{cc}
\epsilon_{1\bk} - i \Gamma_{1\bk} & \Delta_\bk \\
\Delta^*_\bk & \epsilon_{2\bk} - i \Gamma_{2\bk}
\end{array}
\right).
\eeq
The condition for exceptional point  is then given by:
\beq
\epsilon_{1\bK} =\epsilon_{2\bK}, \; |\Delta_{\bK}| = |\Gamma_{1\bK} - \Gamma_{2\bK}|/2.
\eeq
The above two equations can be satisfied at isolated momenta in two-dimensional $\bk$ space, and at lines of momenta in three-dimensions. Indeed, a general phase diagram of Dirac Hamiltonian with both Dirac mass and two-lifetimes consists of an extended region of exceptional points \cite{ShenZhenFu}.
We thus expect that exceptional points are generally present in the spectra of finite-lifetime quasiparticles in two- and three-dimensional systems whose band gap is small or comparable to inverse lifetimes.

\begin{figure}[t]
\centering
\includegraphics[width=0.5\textwidth]{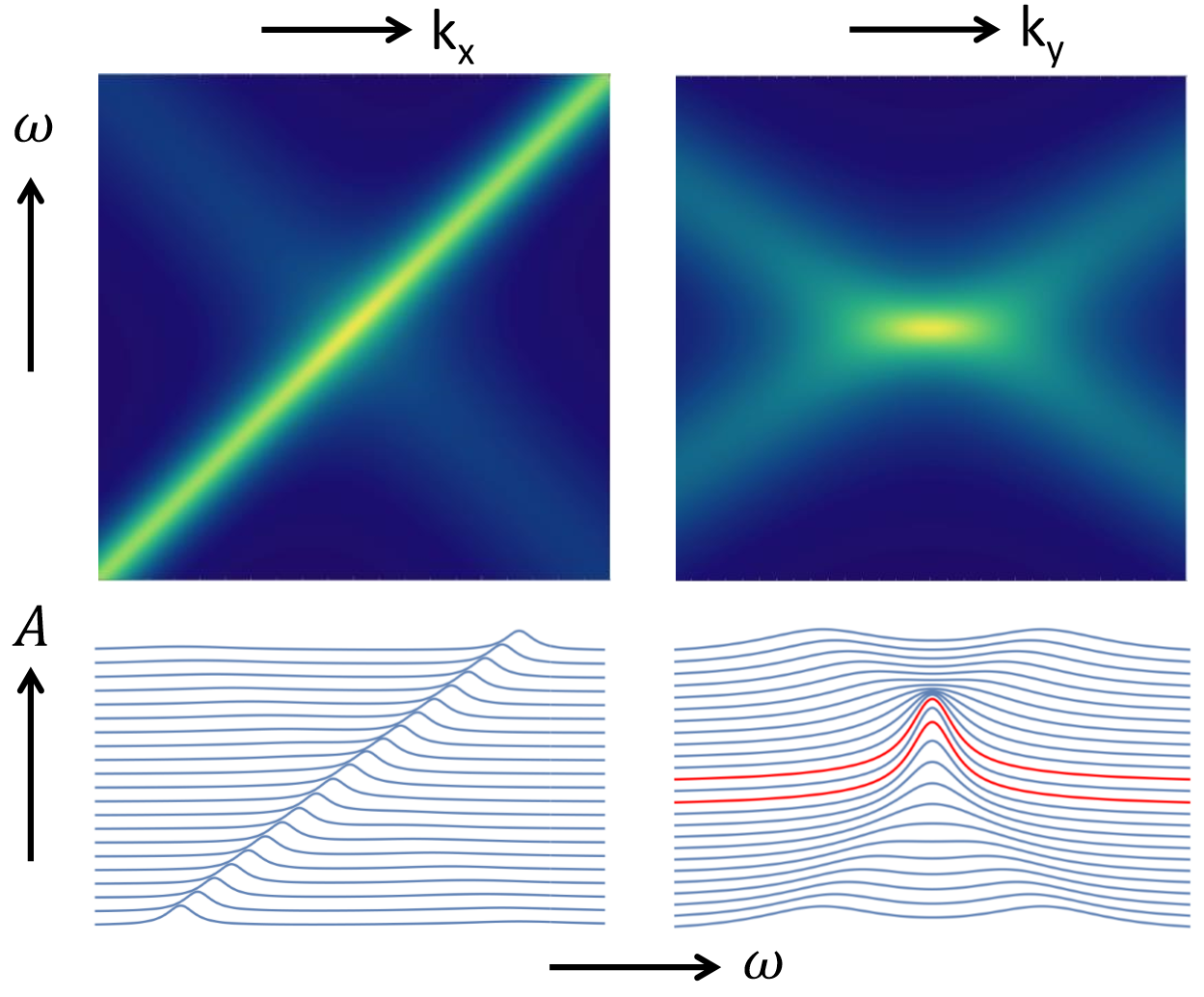}
 \caption{Upper: color intensity map of spectral function $A(\bk, \omega)$ in the $k_x-\omega$ (left) and $k_y-\omega$ (right) plane, showing that quasiparticle conduction and valence bands cross at $k_x=0$ along the $k_x$ axis, but stick over a range of momenta along the $k_y$ axis. Lower: $A(\bk,\omega)$ as a function of $\omega$ for a set of momenta on the $k_x$ (left) and $k_y$ (right) axis, including two exceptional points (in red).}
\end{figure}


The unusual Fermi arc due to self-energy with two lifetimes can be directly probed by angle-resolved photoemission spectroscopy (ARPES), which measures electron spectral function $A(\bk,\omega)= -{\rm Im Tr} ( G^R - G^A)$,
with $G^A \equiv (G^R)^\dagger$. It is straightforward to show that in our model,
\beq
 A(\bk,\omega)= -2 {\rm Im}  ( \frac{1}{\omega - {\cal E}_+(\bk)} + \frac{1}{\omega - {\cal E}_- (\bk)} )
\eeq
is a sum of two Lorentzians centered at the quasiparticle energy ${\rm Re} [{\cal E}_\pm (\bk)]$, whose height and width are determined by the inverse lifetime  ${\rm Im} [{\cal E}_\pm(\bk)]$.
For $\Gamma_1 \neq \Gamma_2$, along the $k_x$ axis two bands  with different spectral weights cross at $k_x=0$, while a line of band degeneracy appears on the $k_y$ axis. This directly demonstrates that the original Dirac band spreads into a bulk Fermi arc, which owes its existence to topological exceptional points.


\begin{figure}[t]
\centering
\includegraphics[width=0.45\textwidth]{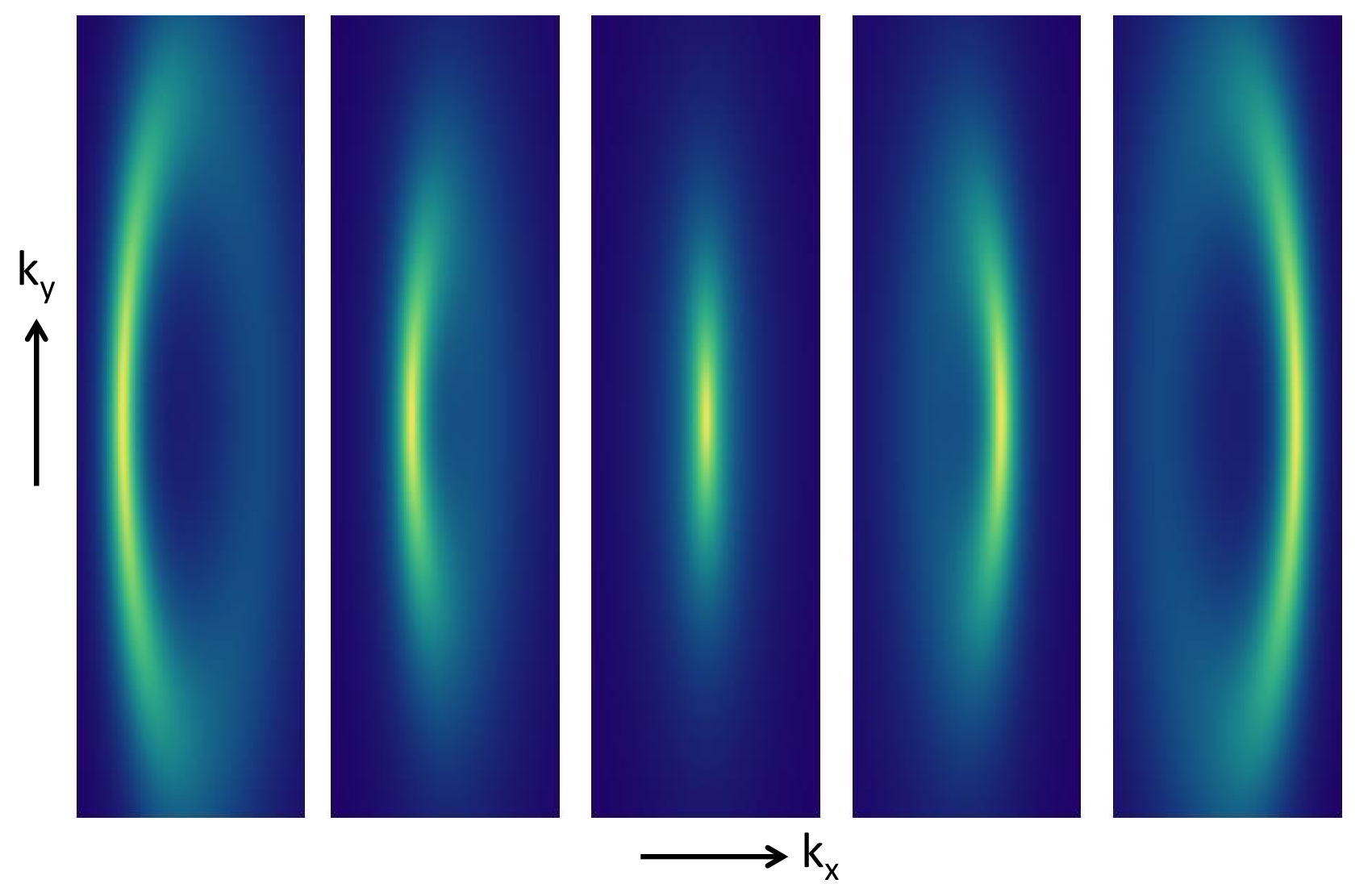}
 \caption{Color intensity map of spectral function in $\bk$ space at constant energy $\omega$, showing the bulk Fermi arc at $\omega=0$ (middle) and  Fermi surfaces with $k_x$-asymmetric spectral weight at $\omega<0$ (left) and $\omega>0$ (right). }
\end{figure}

In addition to creating the Fermi arc at $\omega=0$, the orbital-dependent lifetime also qualitatively changes the spectral function of Dirac semimetals at $\omega \neq 0$. Fig.3 shows an intensity map of spectral function at constant $\omega$ in momentum space. We find that for $\omega \neq 0$, $A(\bk, \omega)$ at $k_x$ and $-k_x$ is asymmetric, such that one side of the Fermi surface is brighter than the other. Moreover, this asymmetry is opposite for energies above and below the Fermi arc. This unusual feature results from the different lifetimes of Dirac quasiparticles at $k_x$ and $-k_x$ having different orbital characters.

To locate the exceptional point requires a more detailed analysis of the lineshape of the spectral function. At the exceptional point, due to the coalescence of eigenvalues ${\cal E}_+={\cal E}_-$,  the spectral function is a {\it single} Lorentzian with a width of $\Gamma=(\Gamma_1+\Gamma_2)/2$. In Fig.2, we plot linecuts of $A(\bk, \omega)$ at two sets of momenta on the $k_x$ and $k_y$ axis respectively, including the exceptional points $\pm \bK=(0,\pm k_0)$. While the Fermi arc defined in terms of the quasiparticle energy ${\rm Re} [{\cal E}_+]= {\rm Re} [{\cal E}_-]$ ends at the exceptional points, the change of spectral function on the $k_y$ axis from having a single peak to two peaks occurs at a momentum $k_y^*$ beyond the exceptional points, i.e., $|k^*_y|>k_0$, because two Lorentzians merge into a single peak when the broadening is larger than the peak separation.

To summarize, our work reveals the dramatic feedback effect of quasiparticle lifetime on quasiparticle energy dispersion in zero- and small-gap semiconductors, including the emergence of topological exceptional points that result in bulk Fermi arcs. The unusual dispersion of these ``exceptional quasiparticles'' can be directly observed by momentum-resolved spectroscopy. They also have strong impact on thermodynamic and transport properties, which we will present elsewhere \cite{ShenFu}. The essential role of self-energy calls for advanced numerical method such as dynamical mean field theory \cite{DMFT} to study and possibly predict real materials that host topological quasiparticles.

Our work shows the usefulness of topology in studying quasiparticles at finite temperature, even when the concept of topological phases cannot be defined.
We hope the transition from topology of quantum phases to topology of finite-lifetime quasiparticles brings  new insight into and a unified understanding of a variety of many-body systems.


\begin{acknowledgements}
LF thanks Bo Zhen for interesting discussions on exceptional points in photonic crystals, which partly inspired this work.
We also thank Sung-Sik Lee, Andrey Chubukov and Jan Zaanen for valuable feedbacks and comments during the KITP conference ``Order, Fluctuations, and Strong Correlations: New Platforms and Developments''.
\end{acknowledgements}


\begin{thebibliography}{1}

\bibitem{TKNN}
D. J. Thouless, M. Kohmoto, M. P. Nightingale, and M. den Nijs
Phys. Rev. Lett. 49, 405 (1982).

\bibitem{HasanKane}
M. Z. Hasan and C. L. Kane, Rev. Mod. Phys. 82, 3045 (2010).


\bibitem{Ando}
Y. Ando, J. Phys. Soc. Jpn. 82, 102001 (2013).

\bibitem{Kato}
T. Kato, Perturbation theory of linear operators (Springer, Berlin, 1966).



\bibitem{ShenZhenFu}
H. Shen, B. Zhen and L. Fu, arXiv:1706.07435

\bibitem{EP-review}
H. Eleuch and I. Rotter, Phys. Rev. A {\bf 95}, 022117 (2017).

\bibitem{BoZhen}
Bo Zhen, Chia Wei Hsu, Yuichi Igarashi, Ling Lu, Ido Kaminer, Adi Pick, Song-Liang Chua, John D. Joannopoulos and Marin Soljacic, Nature, {\bf 525}, 354 (2015).


\bibitem{Coleman}
M. Dzero, J. Xia, V. Galitski, and P. Coleman, Annual Review of Condensed Matter Physics, {\bf 7}, 249 (2016).

\bibitem{Qian}
X. Qian, J. Liu, L. Fu, and J. Li, Science {\bf 346}, 1344 (2014).

\bibitem{TCI-Madhavan}
Ilija Zeljkovic, Yoshinori Okada, Cheng-Yi Huang, R. Sankar, Daniel Walkup, Wenwen Zhou, Maksym Serbyn, Fangcheng Chou, Wei-Feng Tsai, Hsin Lin, A. Bansil, Liang Fu, M. Zahid Hasan, and Vidya Madhavan, Nature Physics {\bf 10}, 572 (2014).

\bibitem{ShenFu}
H. Shen and L. Fu, to appear

\bibitem{DMFT}
S. Y. Savrasov and G. Kotliar, Phys. Rev. B 69, 245101 (2004).



\end{thebibliography}
\end{document}